\documentclass[10pt,aps,prb,twocolumn,amsmath,amssymb,superscriptaddress,showpacs,footinbib]{revtex4-1}
\usepackage{amsmath}
\usepackage{graphicx}
\usepackage{xcolor}
\usepackage{pdfpages}

\begin{document}

\title{Double-resonant LA phonon scattering in defective graphene and carbon nanotubes}%

\author{Felix Herziger}%
\email{fhz@physik.tu-berlin.de}%
\affiliation{Institut f\"ur Festk\"orperphysik, Technische Universit\"at Berlin, Hardenbergstr. 36, 10623 Berlin, Germany}%

\author{Christoph Tyborski}
\affiliation{Institut f\"ur Festk\"orperphysik, Technische Universit\"at Berlin, Hardenbergstr. 36, 10623 Berlin, Germany}%

\author{Oliver Ochedowski}
\affiliation{Fakult\"at f\"ur Physik und CENIDE, Universit\"at Duisburg-Essen, 47048 Duisburg, Germany}

\author{Marika Schleberger}
\affiliation{Fakult\"at f\"ur Physik und CENIDE, Universit\"at Duisburg-Essen, 47048 Duisburg, Germany}

\author{Janina Maultzsch}%
\affiliation{Institut f\"ur Festk\"orperphysik, Technische Universit\"at Berlin, Hardenbergstr. 36, 10623 Berlin, Germany}%

\date{\today}%

\begin{abstract}
We present measurements of the $D''$ Raman mode in graphene and carbon nanotubes at different laser excitation energies. The Raman mode around 1050 - 1150\,cm$^{-1}$ originates from a double-resonant scattering process of longitudinal acoustic (LA) phonons with defects. We investigate its dependence on laser excitation energy, on the number of graphene layers and on the carbon nanotube diameter. We assign this Raman mode to so-called 'inner' processes with resonant phonons mainly from the $\Gamma-K$ high-symmetry direction. The asymmetry of the $D''$ mode is explained by additional contributions from phonons next to the $\Gamma-K$ line. Our results demonstrate the importance of inner contributions in the double-resonance scattering process and add a fast method to investigate acoustic phonons in graphene and carbon nanotubes by optical spectroscopy.
\end{abstract}


\maketitle%

\section{Introduction}
Raman spectroscopy and fundamental research on graphitic systems have been two inherently connected fields for many decades in solid-state physics. Since the early works of Tuinstra and Koenig \cite{10.1063/1.1674108}, Raman spectroscopy has evolved as a very versatile tool to characterize graphite, graphene, and carbon nanotubes \cite{Reich2004, 10.1098/rsta.2004.1454, 10.1098/rsta.2004.1444, 10.1038/nnano.2013.46}. Especially after the introduction of the double-resonance Raman concepts by Thomsen and Reich \cite{PhysRevLett.85.5214}, many phenomena could be unraveled such as the anomalous dispersion of the $D$ and $2D$ modes with laser excitation energy or the origin of many other, higher-order Raman modes. However, most of the works focus on the double-resonant Raman modes related to transverse and longitudinal optical phonons (TO and LO), such as the well-known $D$, $D'$, and the $2D$ mode, or the first-order $G$ band in graphene and the radial-breathing mode (RBM) in carbon nanotubes \cite{PhysRevLett.93.177401, PhysRevB.72.205438, 10.1021/nn2044356, PhysRevB.61.14095, 10.1038/nmat1846, 10.1021/nl061702a, 10.1016/j.carbon.2009.12.057, PhysRevB.82.201409, PhysRevB.79.205433}. Recently, also the investigation of small interlayer vibrational modes in few-layer graphene came into the focus of research, since these modes can directly probe the layer number and interlayer interaction \cite{10.1038/nmat3245, PhysRevB.85.235447, 10.1021/nl302450s, PhysRevB.87.121404}. Beside these modes, there are various other Raman bands that result from double-resonant two-phonon scattering processes or from phonon-defect scattering \cite{PhysRevB.84.035433}. 

In this work, we focus on the defect-induced $D''$ mode (around 1100\,cm$^{-1}$) in graphene and carbon nanotubes that results from double-resonant scattering of longitudinal acoustic (LA) phonons with defects. Low-energy acoustic phonons that stem from the Brillouin-zone center strongly affect charge carrier mobilities and thermal transport properties and are thus very important for the performance of electronic devices \cite{10.1021/nl0731872, PhysRevB.80.195423, 10.1038/nnano.2008.58}. However, first-order Raman scattering does not allow to probe these phonons. The LA phonon was often observed in combination with other phonons in double-resonant Raman scattering processes \cite{PhysRevB.87.075402, 10.1016/j.ssc.2011.05.017}. Especially in bi- and fewlayer graphene, combination modes containing the LA phonon can be observed in the frequency range between the $G$ and the $2D$ mode \cite{PhysRevB.84.035419, 10.1021/nn1031017, 10.1021/nn200010m, PhysRevB.86.195434}. However, none of the previous works analyzed the LA phonon itself. Here, we report the first measurements of the theoretically predicted $D''$ mode in carbon nanotubes and graphene with intentionally created defects \cite{PhysRevB.84.035433}. We investigate its dependence on laser excitation energy, the number of graphene layers and the carbon nanotube diameter. We prove that this Raman mode stems from so-called 'inner' processes, again highlighting the importance of these contributions.

\section{Experimental details}
Graphene samples were prepared by micro-mechanical exfoliation of natural graphite crystals onto silicon substrates with an 100\,nm oxide layer. The layer number of the prepared samples was unambiguously identified by their optical contrast and layer-number dependent Raman modes \cite{PhysRevB.85.235447, 10.1021/nl302450s}. The graphene samples were then transferred to a vacuum chamber and irradiated with swift heavy ions (Xe$^{26+}$, 91 MeV) using a fluence of 65.000 ions/$\mu m^2$ at normal angle to the graphene plane. Under grazing-incidence irradiation these projectiles cause extended modifications in graphene \cite{10.1063/1.3559619, 10.1063/1.4801973}, while at normal incidence point-like defects are created \cite{10.1063/1.4808460}. Due to their high energy, the interaction of swift heavy ions with matter is exclusively by inelastic scattering. As the penetration depth of 91\,MeV Xe ions is about 10\,$\mu$m \cite{10.1016/j.nimb.2010.02.091}, defects are not introduced exclusively in single-layer but in bi- and tri-layer graphene as well. From the fluence of the Xe ions, we deduce an average length between defects of approximately 4\,nm. A determination of the defect length $L_{\text{D}}$ by the $D$/$G$-mode intensity ratio and formulas from Ref. \onlinecite{10.1021/nl201432g} is misleading in the present case. The irradiation with 91\,MeV Xe ions creates significantly smaller defects compared to average defects sizes from previous studies with low-energy Argon ions \cite{10.1016/j.carbon.2009.12.057, 10.1021/nl201432g}. Therefore, we observe a reduced $D$/$G$ mode ratio at the same $L_{\text{D}}$ compared to Ref. \onlinecite{10.1021/nl201432g}. Approximate defect sizes in the present study are $r_{\text{S}} = 0.35$\,nm and $r_{\text{A}} = 2.11$\,nm (using the notation from Ref. \onlinecite{10.1021/nl201432g}). For the measurements on carbon nanotubes, we used buckypaper carbon nanotubes produced by the HiPCO process \cite{10.1016/S0009-2614(99)01029-5}, having a diameter distribution of $10\pm2\,\text{\AA}$, which was verified by measurements of the RBM.

Raman measurements were done with a Horiba HR800 and a Dilor XY spectrometer, equipped with solid-state lasers, as well as dye and gas lasers. Raman spectra were recorded in back-scattering geometry under ambient conditions using a 1800\,lines/mm grating and an $100\times$ objective, yielding a spectral resolution of approx. 1\,cm$^{-1}$. Since the $D''$ mode and the second-order Raman modes of silicon are close in frequency, it was necessary to perform a background subtraction for graphene measured on silicon substrates. For this purpose, we used the same experimental conditions as for the Raman measurements on graphene, \textit{i.e.}, the same laser power and integration time, and recorded the silicon background at a spot adjacent to the investigated graphene flake. During all measurements the laser power was kept below 1\,mW in order to avoid sample heating, laser-induced doping or the creation of additional unwanted defects. 

\begin{figure}
\includegraphics{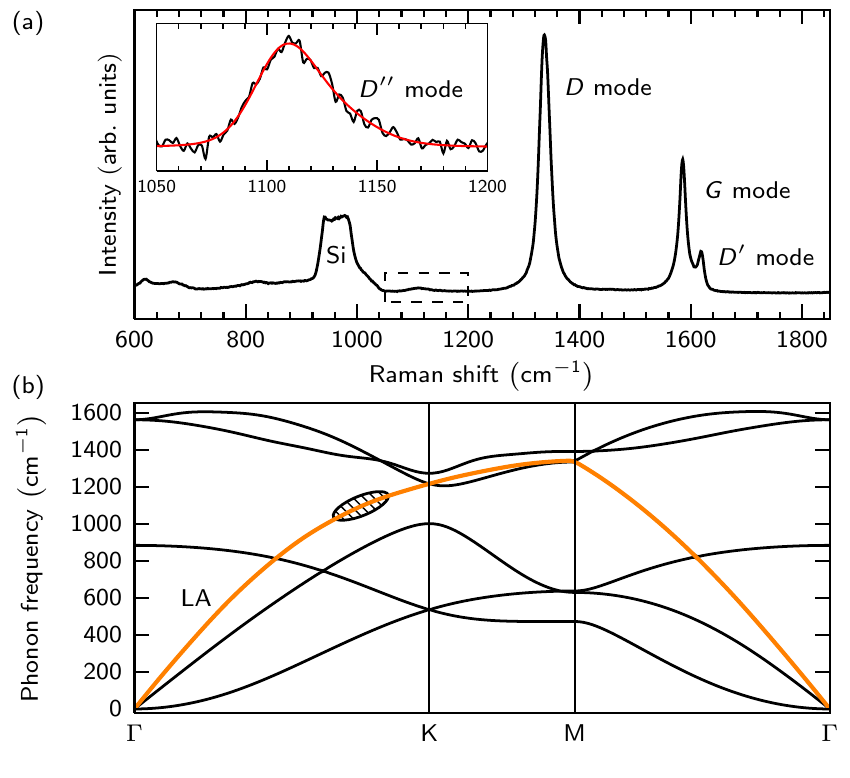}
\caption{(a) Overview spectrum of defective graphene at 532\,nm laser wavelength. The $D$, $G$, $D'$ modes of graphene, as well as the second-order Raman peak of the silicon substrate are labeled. The inset shows an enlarged view of the $D''$ mode at approximately 1100\,cm$^{-1}$. (b) Phonon dispersion of single-layer graphene obtained from \textit{ab-initio} DFT calculation. The LA phonon branch is marked in orange. The hatched ellipse indicates the \textit{k}-space region where the resonant phonons stem from. \label{fig1}}
\end{figure}

\begin{figure*}
\includegraphics{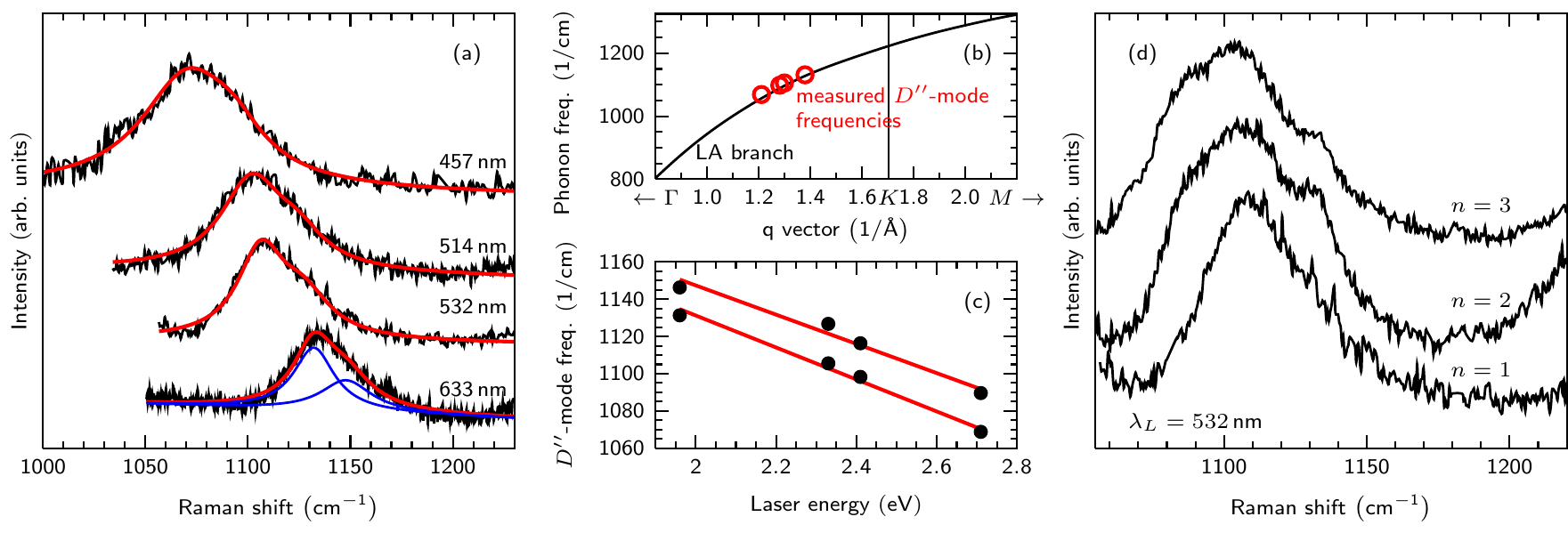}
\caption{(a) Raman spectra of the $D''$ mode in single-layer graphene for different laser excitation wavelengths. The thick solid lines denote fits to the measured spectra. For the spectra recorded at 633\,nm laser wavelength, we additionally show both Lorentzian contributions of the $D''$ mode. (b) Comparison of the measured $D''$-mode frequencies and the LA phonon dispersion along $\Gamma-K-M$. (c) Dispersion of the two $D''$-mode contributions with laser excitation energy. The solid line denotes a linear fit to the data points with a slope of approximately $-80\,\text{cm}^{-1}/\text{eV}$. (d) $D''$-mode spectra at 532\,nm laser wavelength for single-, bi-, and trilayer graphene. The broadening of the $D''$ mode in bi- and trilayer graphene can be clearly seen. \label{fig2}}
\end{figure*}

\section{Results and discussion}
Figure~\ref{fig1}\,(a) shows an overview Raman spectrum of defective single-layer graphene at 532\,nm laser excitation wavelength. The well-known $D$, $G$, and $D'$ Raman modes of single-layer graphene can be clearly identified. By enlarging the spectral range between the second-order Raman peak of the silicon substrate and the $D$ mode, we can identify another Raman mode. We explicitly verified that this Raman mode cannot be observed at arbitrary edges of exfoliated graphene, indicating that a larger number of defects is needed to result in a measurable Raman signal. Following the notation of Venezuela \textit{et al.} in Ref. \onlinecite{PhysRevB.84.035433}, this mode is referred to as $D''$. The observed $D''$/$D$ mode intensity ratio in the spectrum of Fig.~\ref{fig1}\,(a) is approximately 1:100. The small intensity of the $D''$ mode is attributed to the relatively weak electron-phonon coupling of the involved LA phonons compared to the large electron-phonon coupling of the TO-derived phonon modes near the $K$ point in graphene \cite{PhysRevLett.93.185503}. Figure~\ref{fig2}\,(a) presents Raman spectra of the $D''$ mode in single-layer graphene at four different laser excitation wavelengths. The laser-energy dependent peak shift indicates that this Raman mode results from a double-resonance process. The observed downshift of approximately $-80$\,cm$^{-1}$/eV  with increasing laser energy is opposite to the well-known behavior of the $D$ and $2D$ mode in graphene [compare Fig.~\ref{fig2}\,(c)]. This can be explained by the dispersion of the phonon branch that is involved in the scattering process. The $D''$ mode is assigned to a double-resonant intervalley scattering process that involves an LA phonon and a defect, as theoretically predicted in Ref. \onlinecite{PhysRevB.84.035433}. The phonon dispersion of single-layer graphene is shown in Fig.~\ref{fig1}\,(b), the LA branch is highlighted \footnote{\textit{Ab-initio} calculations of phonon dispersions and band structures were performed with the DFT code SIESTA in LDA approximation \cite{10.1088/0953-8984/14/11/302}. Lattice parameters were set to the experimental values from Ref. \onlinecite{PhysRevB.76.035439}. The phonon dispersion was calculated using a $3\times3\times1$ supercell approach.}. Since this Raman mode results from an intervalley scattering process, the dominant contributions stem from phonons along the $\Gamma-K-M$ high-symmetry direction. However, only between $\Gamma$ and $K$ the measured $D''$-mode frequencies match the calculated LA dispersion [compare Fig.~\ref{fig2}\,(b)]. Therefore, we can assign this scattering process to so-called 'inner' processes, \textit{i.e.}, the electronic transition must be between $K$ and $M$. The importance of inner processes was discussed extensively in recent literature on double-resonances in graphene \cite{PhysRevB.84.035433, PhysRevB.87.075402}, showing that the formerly proposed restriction to only 'outer' processes is incorrect \cite{PhysRevLett.97.187401}. Our results furthermore enable us to map the LA phonon branch along the $\Gamma-K$ high-symmetry direction. Figure~\ref{fig2}\,(b) compares the experimentally obtained $D''$-mode frequencies with the calculated LA phonon branch dispersion along $\Gamma-K-M$. The resonant phonon wave-vectors for each excitation energy were obtained from the resonance condition on the incoming and outgoing photon in the double-resonance process; a \textit{GW}-corrected band structure of single-layer graphene was used. The experimental values match the theoretical curve along $\Gamma-K$ within an error of less than 10\,cm$^{-1}$. By tuning the laser excitation energy, one can now follow the LA phonon branch along the high-symmetry line. The experimentally obtained values, including the origin, can be fitted by a sine function; the resulting fit shows a deviation of less than 10\,cm$^{-1}$ close to $\Gamma$ compared to our \textit{ab-initio} calculations (see SI). Thus, our approach enables us to investigate acoustic phonons by an optical method, \textit{i.e.}, double-resonant Raman spectroscopy, not only at certain phonon wave-vectors. In fact, the LA phonon branch, obtained from fitting our experimental data, also showed good agreement close to $\Gamma$, which is the $k$-space region that predominantly affects heat transfer.

We will now turn our discussion to the lineshape of the $D''$ mode. From the Raman spectra in Fig.~\ref{fig2}\,(a) we can derive two statements: First, the $D''$ mode shows a pronounced asymmetry towards higher frequencies and second, the full width at half maximum (FWHM) of $D''$ mode increases with increasing laser excitation energy. The asymmetric tail can be seen very clearly in all spectra. We fitted all Raman spectra with two Lorentzian components; the spectrum at 633\,nm laser wavelength exemplarily shows their individual contributions. The asymmetry of the $D''$ mode can be directly explained with the two-dimensional phonon dispersion of the LA branch around the $K$ point. In contrast to the TO branch, the LA phonon branch shows a non-constant angular frequency dependence around the $K$ point, \textit{i.e.}, the lowest frequencies can be found along $K-\Gamma$ and the highest along $K-M$ \cite{PhysRevB.87.075402}. Thus, we can assign the intense, lower-frequency Lorentzian to contributions directly from the $\Gamma-K$ high-symmetry direction. The high-frequency tail of the $D''$ peak can be assigned to contributions next to the high-symmetry line in the Brillouin zone, following the analysis of May \textit{et al.} \cite{PhysRevB.87.075402}. The broadening of the lineshape at larger excitation energies can be easily explained by considering the dispersion of the LA branch along $\Gamma-K$, which has a steeper dispersion closer to $\Gamma$ [see Fig.~\ref{fig1}\,(b)]. Therefore, for shorter resonant phonon wave-vectors, \textit{i.e.}, larger excitation energies, a larger frequency range can be accessed, leading to a broadening of this Raman mode. Similar observations, both theoretically and experimentally, were made for the $D+D''$ mode at approximately 2450\,cm$^{-1}$ in the Raman spectrum of single-layer graphene \cite{PhysRevB.87.075402}.

The comparison of our measured $D''$-mode spectra with calculations from Venezuela \textit{et al.} (compare Ref. \onlinecite{PhysRevB.84.035433}, Fig.~11) shows very good agreement, both qualitatively and quantitatively. The prediction of an asymmetric high-frequency tail was confirmed by our measurements. Also the absolute frequencies match the experimentally observed, supporting our assignment to the LA-defect scattering process.

Figure~\ref{fig2}\,(d) shows $D''$-mode spectra at 2.33\,eV laser excitation energy for different numbers of graphene layers. As can be seen very clearly, the lineshape significantly broadens when going from single-layer to bilayer graphene. This broadening can be directly attributed to the evolution of the electronic bandstructure around the $K$ point. Since bilayer graphene has two valence and conduction bands, the number of resonant scattering processes is quadrupled compared to single-layer graphene. The increased number of resonant phonon wave-vectors leads to an increased number of resonantly enhanced phonons, thus resulting in a broadening of the Raman mode. This effect can be also observed for the $D+D''$ mode in bilayer graphene \cite{PhysRevB.87.075402}, as well as for the $2D$ mode \cite{PhysRevLett.97.187401}. By further increasing the layer number, the lineshape does not show any noticeable changes in linewidth or by the appearance of additional peaks. This can be again identified with the shape of the electronic bands around $K$. The bandstructure of trilayer graphene can be regarded as a superposition of the electronic bands of single and bilayer graphene \cite{10.1073/pnas.1004595107}. Thus, also the resonant phonon wave-vectors in the double-resonance process are very similar to the ones from single and bilayer graphene. Therefore, the $D''$-mode lineshape in trilayer graphene does not differ from the peak observed in bilayer graphene. Due to the degeneracy of the LA phonon branch in bi- and tri-layer graphene \cite{PhysRevB.77.125401}, the broadening of the $D''$ mode cannot result from the evolution of the phonon dispersion with increasing layer number, but rather from the evolution of the electronic bandstructure.

Similar to graphene, only a few publications reported double-resonant Raman modes in carbon nanotubes that involve LA phonons \cite{PhysRevB.66.245410, PhysRevB.66.155418, J.Phys.ChemC112, AIP125}. Besides intravalley scattering with LO and LA derived phonons, also intervalley scattering with TO and LA derived phonons was reported \cite{PhysRevB.66.155418, 10.1016/j.carbon.2004.11.044, 10.1021/jp803855z, 10.1063/1.2354081}. However, double-resonance processes combining a defect and an LA phonon have not been reported so far in carbon nanotubes.

\begin{figure}
\includegraphics{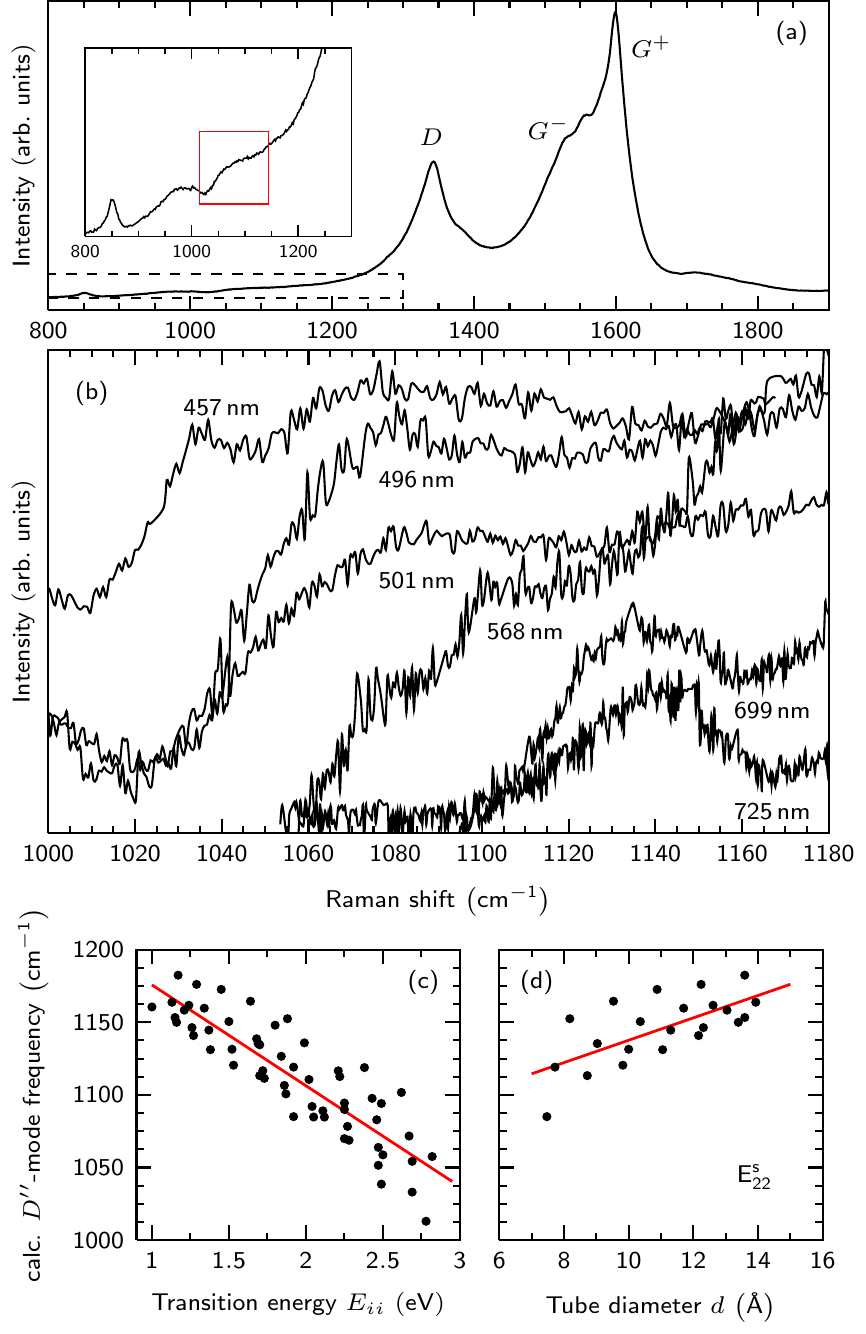}
\caption{(a) Overview Raman spectrum of HiPCO carbon nanotubes at 514\,nm laser wavelength. The inset shows an enlarged view of the $D''$-mode region; the $D''$mode is marked with the red rectangle. (b) Raman spectra of the $D''$ mode in carbon nanotubes for various laser excitation energies. Spectra are vertically offset for clarity. (c) Calculated $D''$-mode frequencies for carbon nanotubes in the diameter range between 7\,\AA{} and 14\,\AA. (d) Calculated frequencies as a function of the tube diameter for the semiconducting $E_{22}$ transition. The solid lines in (c) and (d) represent linear fits to the calculated data points. \label{fig3}}
\end{figure}

Figure~\ref{fig3}\,(b) shows Raman spectra of the $D''$ mode in carbon nanotubes for six different laser excitation energies. Due to a broader $D$ mode in carbon nanotubes compared to graphene \cite{10.1002/pssb.201200175}, the $D''$ mode can only be observed as a low-frequency shoulder to the $D$ mode [compare Fig.~\ref{fig3}\,(a)]. From the experimental spectra we estimate a downshift of the $D''$ mode with higher laser excitation energies of approximately $-75$\,cm$^{-1}$/eV. In contrast to graphene, we observe a complex peak structure with many different contributions. This can be understood from the fact that the Raman spectrum of carbon nanotubes results from many different tubes in or close to resonance with the excitation laser. The analyzed HiPCO sample contains a large variety of carbon nanotubes with diameters around $10\,\text{\AA}$. Therefore, the $D''$-mode lineshape is broadened by the different contributions of carbon nanotubes in or close to resonance. However, an assignment of the different features in the $D''$ band to distinct carbon nanotubes is not possible because of two reasons: the large number of different carbon nanotubes analyzed and the very low intensity of the $D''$ band. 

In order to derive a systematic analysis, we present calculated $D''$-mode frequencies for all carbon nanotubes in the diameter range between 7\,\AA{} and 14\,\AA{}. The calculation is based on a sixth-nearest neighbor tight-binding model with symmetry-imposed modifications for carbon nanotubes using the POLSym code \cite{polsym}. The calculated LA phonon branch is scaled in frequency by 5.1\,\% in order to fit the experimentally observed value at the $K$ point for graphite \cite{PhysRevB.80.085423}. The resonant phonon wave vectors in the double-resonance process were obtained by assuming that scattering occurs between two equivalent extrema in the electronic band structure. As indicated by other works before \cite{PhysRevB.87.165423, PhysRevB.84.035433, 10.1002/pssb.201200175}, we assumed that the dominant contribution to the double-resonance process results from the incoming resonance. Similar to graphene, the oscillator strength for optical transitions in carbon nanotubes is highest along the wavevectors derived from the $K-M$ direction in the graphene Brillouin zone \cite{PhysRevB.74.195431, PhysRevB.74.115415}. Therefore, the Raman spectrum is dominated by carbon nanotubes, where these transitions are probed, \textit{e.g.}, the $\nu=-1$ family for the semiconducting $E_{22}$ transition \footnote{The family index $\nu$ of a carbon nanotube is given by $\nu = (n_1 - n_2)\mod3$, where $n_1$ and $n_2$ are the chiral indices of the CNT.}. Thus, we restrict our calculations to those transitions. The calculated $D''$-mode frequencies in Fig.~\ref{fig3}\,(c) reproduce the experimentally observed peak positions and peak shift very well. The calculated shift rate of $-70$\,cm$^{-1}$/eV is in reasonable agreement with the experiments. As can be seen in Fig.~\ref{fig3}\,(c), the calculated $D''$-mode frequencies at each laser energy cover a large frequency range of approximately 40\,cm$^{-1}$, in accordance with the experimentally observed broad lineshape of the $D''$ mode in carbon nanotubes. Fig.~\ref{fig3}\,(d) shows the calculated frequencies as a function of the tube diameter for the semiconducting $E_{22}$ transition (all other transitions show the same behavior). We observe an upshift of approximately 7\,cm$^{-1}$/\AA{} with increasing tube diameter. Since the LA phonon branch itself shows nearly no dependence on the tube diameter \cite{PhysRevB.75.035427, polsym}, the observed diameter dependence basically reflects the diameter dependence of the optical transition energies.

\section{Conclusion}
In summary, we presented experimental Raman spectra of the $D''$ mode in graphene and carbon nanotubes at various laser excitation energies. We showed that this mode results from double-resonant intervalley scattering of LA phonons with defects and has a dispersion of $-80\,\text{cm}^{-1}/\text{eV}$ in single-layer graphene. We demonstrated that the $D''$ must stem from so-called 'inner' scattering processes with additional contributions from phonons next to the $\Gamma-K$ direction, explaining the observed high-frequency tail of this Raman mode. We further showed that the lineshape in graphene depends on the layer number, reflecting the evolution of the electronic bands around $K$. In carbon nanotubes, the lineshape of the $D''$ mode is significantly broadened due to contributions from different tubes in or close to resonance with excitation laser. Our theoretical calculations of this Raman mode in carbon nanotubes showed very good agreement with the experimental data. Our analysis presents a fast and simple method to investigate acoustic phonon branches by an optical method. Although the probed LA phonons stem from a region close to the $K$ point, we demonstrated the possibility to derive the LA dispersion along the complete $\Gamma-K$ direction. Furthermore, our results again highlight the importance of inner contributions to the double-resonance Raman process in graphitic systems. 

\appendix
\begin{acknowledgments}
The authors acknowledge H. Lebius and B. Ban-d'Etat from the IRRSUD beamline of the Grand Accelerateur National d'Ions Lourds GANIL, Caen, France, for performing the irradiation of the exfoliated graphene samples. This work was supported by the European Research Council (ERC) under grant no. 259286 and by the DFG under grant number MA 4079/7-2 and  SCHL 384/15-1 within the SPP1459 ``Graphene''. 
\end{acknowledgments}

\bibliographystyle{apsrev4-1}

\newpage
\includepdf{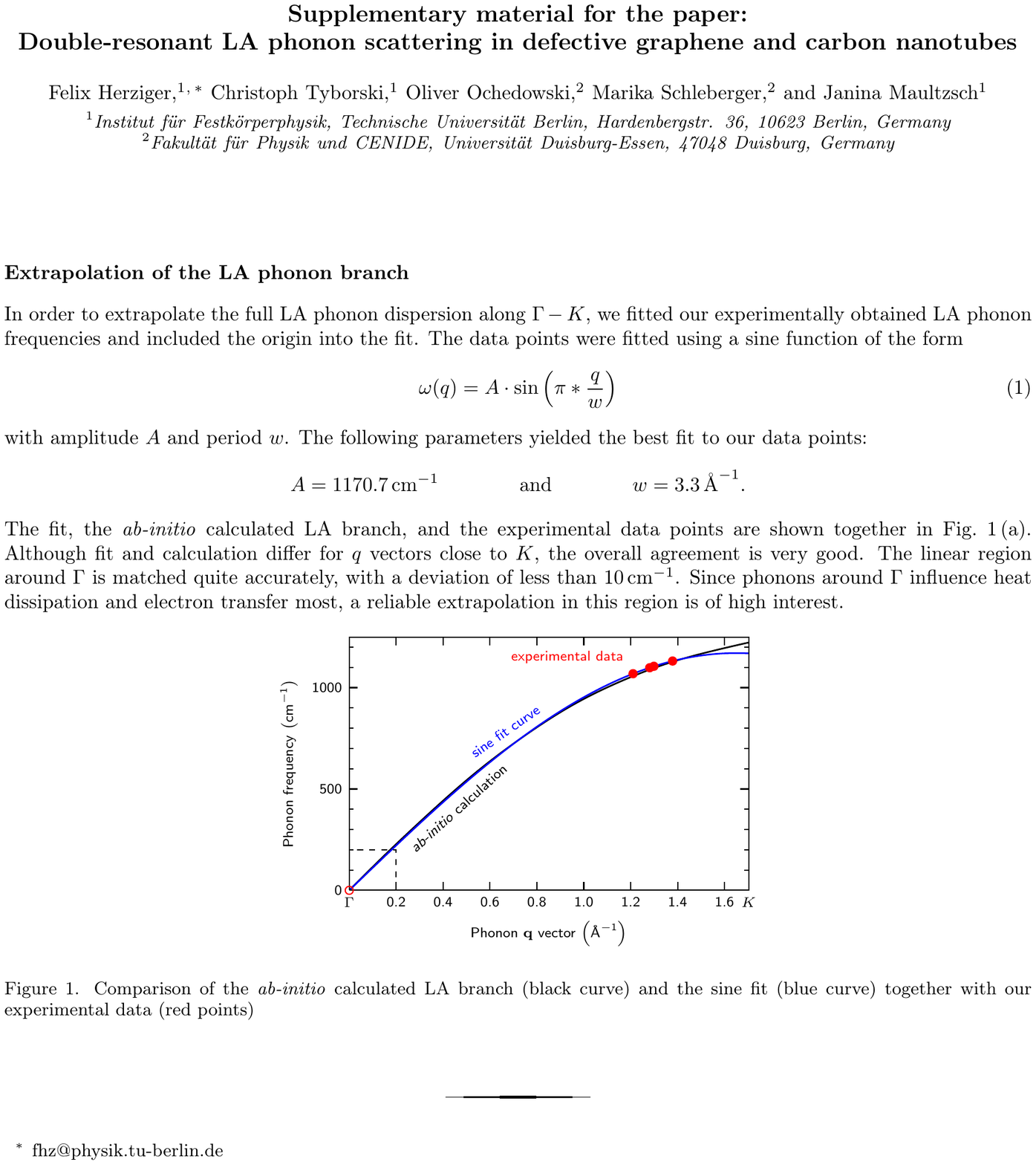}

\end{document}